\documentclass{ws-ijmpd}
\usepackage[super,compress]{cite}

\usepackage{amssymb}
\usepackage{color}
\usepackage{amsmath}
\usepackage{amsfonts}
\usepackage{verbatim}

\newcommand{\db}{\boldsymbol{\delta}_B}
\newcommand{\dl}{\delta \lambda}
\newcommand{\dph}{\delta \phi}
\newcommand{\dpt}{\widetilde{\delta \phi}}
\newcommand{\ul}{u_\lambda}
\newcommand{\up}{u_\phi}
\newcommand{\pl}{\pi_\lambda}
\newcommand{\pp}{\pi_\phi}
\newcommand{\pb}{\pi_b}
\newcommand{\pc}{\pi_c}
\newcommand{\bo}[1]{\textbf{#1}}

\begin{document}

\markboth{Rio Saitou, Yungui Gong}
{de Sitter spacetime with a Becchi-Rouet-Stora quartet}

%
\catchline{}{}{}{}{}
%

\title{de Sitter spacetime with a Becchi-Rouet-Stora quartet}

\author{Rio Saitou}

\address{School of Physics, Huazhong University of Science and Technology,  \\
Wuhan 430074, China
\\
riosaitou@hust.edu.cn}

\author{Yungui Gong}

\address{School of Physics, Huazhong University of Science and Technology,  \\
Wuhan 430074, China
\\
yggong@hust.edu.cn}

\maketitle


\begin{abstract}
We generalize the topological model recently proposed and investigate the cosmological perturbations of the model.
The model has an exact de Sitter background solution associated with a Becchi-Rouet-Stora(BRS) quartet terms which are regarded as a Lagrangian density of the topological field theory. The de Sitter solution can be selected
without spontaneously breaking the BRS symmetry, and be interpreted as a gauge fixing of de Sitter spacetime. 
The BRS symmetry is preserved for the perturbations around the de Sitter background before we solve the constraints of general relativity.
We derive action to the second order of the perturbations and confirm that
even after solving the constraints, we have the BRS symmetry at least for the second order action.
We construct the cosmological perturbation theory involving the BRS sector,
and obtain the two point correlation functions for the curvature perturbation and the isocurvature perturbations which compose the BRS sector.
Our result gives a new description for de Sitter spacetime and the quantum field theory in de Sitter spacetime.
\end{abstract}


\ccode{PACS numbers: 95.36.+x, 11.90.+t}

\section{Introduction}

It has been confirmed by several astronomical observations \cite{Perlmutter:1998np, Riess:1998cb, Eisenstein:2005su, Hinshaw:2012aka, Ade:2015xua} that the
universe is currently accelerating driven by an unknown energy, called the dark energy. Among the candidates of dark energy,
the cosmological constant is the oldest and probably the most accepted candidate, although it has some serious problems
about naturalness \cite{Zeldovich:1967gd,Weinberg:1988cp}.
Especially, to explain the present cosmic acceleration, the energy scale of the cosmological constant must be fine tuned
to be extremely smaller than any typical energy scale of the vacuum energy expected from the quantum field theory including the gravity.
However, there is no convinced reason for the smallness of the cosmological constant in the
quantum field theory.

To tackle this fine-tuning problem, the models called unimodular gravity \cite{Anderson:1971pn, Buchmuller:1988wx, Buchmuller:1988yn, Henneaux:1989zc, Unruh:1988in, Ng:1990xz}
and the models which have ``sequestering" mechanism \cite{Kaloper:2013zca, Kaloper:2014dqa, Kaloper:2015jra}
were suggested. These models can sequester the cosmological constant or the vacuum energy from the evolution
of the universe by imposing constraints on the measure or the cosmological constant.

Motivated by the unimodular gravity, the topological model was proposed recently \cite{Nojiri:2016mlb}.
This model has the Einstein-Hilbert term with a cosmological constant, and a vanishing Lagrangian density for the scalar field $\phi$, ${\cal L}_\phi=0$, so that the theory trivially has a ``gauge symmetry" under any transformations for $\phi$. As gauge fixing terms, the Becchi-Rouet-Stora(BRS) quartet terms were introduced.
By redefinition of one of the scalar fields in the BRS sector, it is argued that we could sequester the cosmological constant from the evolution of the universe.
Surprisingly, although the BRS sector is the unphysical mode which can not be observed,
one of the scalar fields in the BRS sector can have nonzero background value in the de Sitter(dS) spacetime.
We could interpret the model as a topological field theory \cite{Witten:1988ze}, where the Lagrangian density is given by the BRS exact form.\footnote{There are other models of the topological field theory in the context of the induced gravity \cite{Oda:2016tdr}.}

In this work, we generalize the topological model so as to have a generic potential of a scalar field $\lambda$, which is one of the BRS sector, and derive a dS solution for the model. We give an interpretation to the dS solution and the sequestering of cosmological constant by considering a Ward-Takahashi(WT) identity. Then, 
we investigate perturbations around the dS background in the topological model, and aim to construct the cosmological perturbation theory \cite{Mukhanov:1990me} using the BRS sector. 
The perturbations in the topological model have not been studied yet, and it is quite nontrivial
to study how they behave around the dS background.
As a major premise, we must keep the BRS symmetry also for the perturbations without the spontaneous symmetry breaking so as to eliminate the negative norm states from the physical Fock space of the theory. Unlike the Minkowski spacetime, however, the perturbations around the dS background
interact nonlinearly with the gravity sector which posesses constraints. It is uncertain
whether we have the BRS symmetry for the perturbations after taking the dS background and dealing with the gravitational constraints.
Moreover, even if we can preserve the BRS symmetry for the perturbations, it is obscure
whether we can have any dynamical perturbation for the BRS sector. 
The original theory is given by ${\cal L}_\phi=0$, so we have the graviton only without adding the BRS sector. Would this situation be changed if we add the BRS quartet terms and expand them around the dS background solution?
These questions motivate us to investigate the BRS symmetry for the perturbations and the field theory around the dS background.

This article is organized as follows. After we shortly review the topological model in the next section, we generalize the model and derive the dS background solution in section \ref{sec31}.
In section \ref{sec3},
we analyze the perturbations around the dS background and discuss about the BRS symmetry for the perturbations. We will naturally assume that we can preserve the BRS symmetry for the perturbations regardless of the gravitational constraints. We also derive the second order action for the perturbations. Under the assumption that we can preserve the BRS symmetry, in section \ref{sec4}, we construct the cosmological perturbation theory and calculate the two point functions of curvature and isocurvature perturbations caused by the linear perturbations in the BRS sector. Section \ref{sec5} is devoted to conclusion.

\section{Topological model}\label{sec2}

We review the topological model proposed in Ref.~\refcite{Nojiri:2016mlb}.
We begin with the following action
\begin{align}
\label{Sp}
    S= \int d^4x\sqrt{-g}\left[ \frac{M_\text{Pl}^2}{2}R - \Lambda - \lambda\left( \mu^3 + \Box \phi\right)\right] + S_m[g_{\mu\nu},\Psi] \ ,
\end{align}
where $M_\text{Pl}$ is the Planck mass, $R$ is the Ricci scalar, $\nabla_\mu$ is the covariant derivative compatible with $g_{\mu\nu}$,
$\Box \equiv g^{\mu\nu}\nabla_\mu\nabla_\nu$, $\mu$ is a parameter which has the mass dimension 1
and $\Lambda$ is the cosmological constant. $\lambda$ and $\phi$ are scalar fields\footnote{We rescale the scalar field $\lambda$ compared with Ref.~\refcite{Nojiri:2016mlb} so that we can define the action (\ref{Sp}) if $\mu=0$.}, and $S_m$ is the action for ordinary matters $\Psi$ such as the Standard Model particles.
If we redefine the field $\lambda$ as $\lambda\rightarrow \lambda - \frac{\Lambda}{\mu^3}$, we can eliminate the cosmological constant as a total derivative
\begin{align}
\label{SS}
    S= \int d^4x\sqrt{-g}\left[ \frac{M_\text{Pl}^2}{2}R - \lambda\left( \mu^3 + \Box \phi\right)\right] + \frac{\Lambda}{\mu^3}\int d^4x \sqrt{-g} \Box \phi + S_m[g_{\mu\nu},\Psi]  \ .
\end{align}
Hence, in Ref.~\refcite{Nojiri:2016mlb}, it is argued that we could drop the cosmological constant from the action and the evolution of the universe. We ignore the total derivative term and use the redefined scalar field $\lambda$ hereafter.

In the action \eqref{SS},
both of the scalar fields are dynamical, but contain a ghost field. In fact, if we change the dynamical  variables as
\begin{align}
\label{}
    \lambda= \frac{1}{\sqrt{2}}(\alpha-\beta)\ , \quad \phi = \frac{1}{\sqrt{2}}(\alpha + \beta)  \ ,
\end{align}
we obtain
\begin{equation}
\label{ }
S = \int d^4x\sqrt{-g}\left[ \frac{M_\text{Pl}^2}{2}R -  \frac{1}{2}g^{\mu\nu}\partial_\mu \alpha\partial_\nu \alpha + \frac{1}{2}g^{\mu\nu}\partial_\mu \beta\partial_\nu \beta - \frac{\mu^3}{\sqrt{2}}(\alpha - \beta) \right] + S_m[g_{\mu\nu},\Psi]\ .
\end{equation}
The field $\beta$ is a ghost since its kinetic term has the wrong sign and creates negative norm states when it is quantized.

To avoid the negative norm states, we add Grassmann odd ghosts $b$ and $c$ to the action (\ref{SS}) as
\begin{align}
\label{S}
    S&= \int d^4x\sqrt{-g}\left[ \frac{M_\text{Pl}^2}{2}R -  \lambda\left( \mu^3 + \Box \phi\right) -i b\Box c\right]  + S_m[g_{\mu\nu},\Psi] \nonumber \\
     &= \int d^4x\sqrt{-g}\left[ \frac{M_\text{Pl}^2}{2}R -  \mu^3\lambda  + g^{\mu\nu}\partial_\mu \lambda\partial_\nu \phi + i g^{\mu\nu}\partial_\mu b\partial_\nu c\right]  + S_m[g_{\mu\nu},\Psi]
\end{align}
We note that the ghosts $b$ and $c$ are scalar fields, but obey the Fermi-Dirac statistics when they are quantized\footnote{We add a factor $i$ to the ghost term which was not given in Ref.~\refcite{Nojiri:2016mlb} to set anti-commutation relations in the usual way \cite{Kugo:1977zq, Kugo:1979gm}.}.
We define a BRS transformation
\begin{equation}
\label{BRS}
\delta_B \lambda = \delta_B c = 0,\quad \delta_B \phi = \epsilon c,\quad  \delta_B b = i\epsilon\left(\lambda - \frac{\Lambda}{\mu^3}\right)\ ,
\end{equation}
where $\epsilon$ is a Grassmann odd parameter.
By removing the parameter $\epsilon$ from the transformation as $\delta_B = \epsilon \db$, we can define the deformed BRS operator $\db$.
$\db$ satisfies nilpotency, $\db^2=0$, and does not act on gravity and ordinary matter sectors, $\db g_{\mu\nu} = \db\Psi = 0$. The action (\ref{S}) is invariant under the BRS transformation by $\db$.
Thus, the scalar fields $\lambda,\ \phi,\ b$ and $c$ compose the BRS quartet. Defining physical states as the states invariant under the BRS transformation, 
we could remove the negative norm states by Kugo-Ojima's quartet mechanism as in the Yang-Mills gauge theory \cite{Kugo:1977zq, Kugo:1979gm}.

The Lagrangian density of scalar fields
\begin{equation}
\label{L}
\mathcal{L} = \sqrt{-g}\left( -\mu^3\lambda + g^{\mu\nu}\partial_\mu \lambda\partial_\nu \phi + i g^{\mu\nu}\partial_\mu b\partial_\nu c\right)
\end{equation}
can be written as a BRS exact form up to total derivative terms
\begin{align}
\label{dL}
    \mathcal{L} =& -i\db\left\{ b\,\sqrt{-g}\left(-\mu^3 - \Box \phi\right)\right\}  \\
     &+ \text{total derivative terms}   \ .\nonumber
\end{align}
If we ignore the total derivative terms, we can regard the theory as a topological field theory \cite{Witten:1988ze} where the Lagrangian density is BRS exact. Let us consider a field theory of $\phi$, and give a vanishing Lagrangian density, $\mathcal{L}_\phi=0$. The theory is trivially invariant under any transformation of $\phi$, and therefore, it can be regarded as a gauge theory.
To fix the gauge, we add a gauge fixing term
\begin{equation}
\mathcal{L} = -i\db (b \,F[\lambda,\,\phi,\,b, c, g_{\mu\nu},\Psi]) \ ,
\end{equation}
where $F$ is a gauge fixing function whose ghost number must be zero.
If we choose the gauge fixing function $F$ as that in (\ref{dL}), we obtain the Lagrangian density \eqref{L} as the BRS quartet terms with $\mathcal{L}_\phi=0$.


Unfortunately, when $\mu=0$, we cannot define the BRS transformation (\ref{BRS}) and the theory as well. In the next section, we propose a new model with a generic potential of $\lambda$, and provide an example which is well-defined anytime.

\section{Topological model with a generic potential of $\lambda$}\label{sec31}

We propose a new topological model by giving another gauge fixing function $F$ which creates a generic potential of $\lambda$. 
If we consider other gauge fixing functions which create more general Lagrangians, we obtain complicated interaction terms in the action, and it becomes difficult to solve the motions of fields.  
On the other hand, we can solve them relatively easily if we generalize the potential of $\lambda$ only. The obtained model have a de Sitter solution.
Then, we give an interpretation of the de Sitter solution of the model.

\subsection{Eliminating $\Lambda$ by a generic potential of $\lambda$}

We consider the following action and gauge fixing function
\begin{align}
\label{Sf}
    S&= \int d^4x\left\{\sqrt{-g}\left( \frac{M_\text{Pl}^2}{2}R - \Lambda \right) -i \db(b \,F[\lambda,\,\phi,\,b, c, g_{\mu\nu},\Psi]) \right\}
    + S_m[g_{\mu\nu},\Psi] \ , \nonumber \\
    F &= \sqrt{-g}(-\Box \phi - V(\lambda)) \ ,
\end{align}
where $V(\lambda)$ is a general function of $\lambda$.
We define the BRS transformation as
\begin{equation}
\label{B2}
\db \lambda = \db c =0, \quad \db \phi =  c,\quad \db b = i(\lambda-M(\Lambda)) \ ,
\end{equation}
where $M(\Lambda)$ is a smooth function of $\Lambda$ for $\Lambda\geq 0$. We assume that there exists a function $N(\Lambda)$ which satisfies 
\begin{equation}
\label{ }
MN = N M = \Lambda \ .
\end{equation}
We set the function $V(\lambda)$ as
\begin{equation}
\label{ }
V = v(\lambda) + N(\Lambda)\ ,
\end{equation}
where $v(\lambda)$ is a smooth function of $\lambda$.
Then, we get the action in which the cosmological constant $\Lambda$ is eliminated  
\begin{eqnarray}
\label{SU}
    &&S= \int d^4x\sqrt{-g}\left[ \frac{M_\text{Pl}^2}{2}R - U(\lambda) + g^{\mu\nu}\partial_\mu \lambda\partial_\nu \phi + i g^{\mu\nu}\partial_\mu b\partial_\nu c\right]  + S_m[g_{\mu\nu},\Psi], \\
    \label{U}
&&U(\lambda) 
                  = (\lambda-M)v + N\lambda \ .
\end{eqnarray}
The potential $U(\lambda)$ is a smooth function of $\lambda$ since $v(\lambda)$ is a smooth function of $\lambda$. In general, $\Lambda$ will appear in the action not as a constant but as coefficients in the potential. Choosing $M = \Lambda/\mu^3$ and $V= \mu^3$, we get the model in the last section. 
If we give the following $M$, $N$ and $v$ 
\begin{align}
\label{ }
M &= {}^4\sqrt{\Lambda},\quad N= {}^4\sqrt{\Lambda^3}, \nonumber \\
 v &= \lambda^3 + {}^4\sqrt{\Lambda}\lambda^2 + \sqrt{\Lambda}\lambda \ ,
\end{align}
we obtain a potential without coefficients
\begin{equation}
\label{ }
U = \lambda^4 \ .
\end{equation}
In this example, we can define the BRS transformation (\ref{B2}) as long as $\Lambda\geq0$, and eliminate $\Lambda$ completely from the action anytime. 

The action (\ref{SU}) is invariant under the BRS transformation (\ref{B2}). 
The corresponding conserved charge is given by
\begin{equation}
\label{QB}
Q_B = \int d^3x\sqrt{-g}g^{0\mu}\left\{ \partial_\mu \lambda c - \partial_\mu c(\lambda-M)\right\} \ .
\end{equation}
The action  is also invariant under the scale transformation $c\rightarrow {\rm e}^\rho c$, $b \rightarrow {\rm e}^{-\rho}b$ where $\rho$ is an arbitrary real number. The corresponding conserved charge
\begin{equation}
\label{QC}
Q_c = -i\int d^3x\sqrt{-g}g^{0\mu}\left(b\,\partial_\mu c - \partial_\mu b \,c\right)
\end{equation}
assigns the ghost number $1$ for $c$ and $-1$ for $b$. 
Defining the physical states using the BRS charge (\ref{QB}) as
\begin{equation}
\label{phys}
Q_B\left.|\text{phys}\right> = 0\ ,
\end{equation}
we could remove the negative norm states from the model by the quartet mechanism.

\subsection{de Sitter solution and its interpretation}

We investigate a cosmological background solution for the action \eqref{SU}. Hereafter, we ignore the matter action $S_m$ for simplicity. We assume the flat Friedmann-Robertson-Walker(FRW) metric
\begin{equation}
\label{ }
ds^2 = -dt^2 + a^2(t)d\bo{x}^2 \ ,
\end{equation}
and homogeneous background fields $\phi = \bar{\phi}(t),\, \lambda =\bar{\lambda}(t)$. We set the background values for the fields $b$ and $c$ to be zero since they have ghost numbers.

The Einstein equations for the backgrounds are
\begin{align}
\label{F}
    3M_\text{Pl}^2H^2&=   U(\bar{\lambda}) - \dot{\bar{\lambda}}\dot{\bar{\phi}} \ ,   \\
    -M_\text{Pl}^2(  3H^2 +2\dot{H})&=  -U(\bar{\lambda}) - \dot{\bar{\lambda}}\dot{\bar{\phi}} \ ,
\end{align}
where the dot denotes the derivative with respect to the time $t$ and $H \equiv \dot{a}/{a}$ is the Hubble parameter.
The field equations for $\bar{\lambda}$ and $\bar{\phi}$ are
\begin{align}
    \label{lambda}
    & \ddot{\bar{\lambda}} + 3H\dot{\bar{\lambda}} = 0 \ , \\
    \label{phi}
    &\ddot{\bar{\phi}} + 3H\dot{\bar{\phi}} - u_1 = 0, \quad u_1\equiv \left. \frac{dU}{d\lambda}\right|_{\lambda=\bar{\lambda}} \ .
\end{align}
The general solution for $\bar{\lambda}$ is given by
\begin{equation}
\label{so}
\bar{\lambda} = C_0 + C_1\int \frac{dt}{a^3(t)} \ ,
\end{equation}
where $C_0$ and $C_1$ are the constants of integration. 
To preserve the BRS symmetry, however, we must determine the background value of $\lambda$ by  the following Ward-Takahashi(WT) identity\footnote{This point was missed in Ref.~\refcite{Nojiri:2016mlb}, and we take it into account for the first time.}
\begin{align}
\label{WT}
   &0 = \left<0|\{iQ_B,\,b\}|0\right>
         = i\left<0|(\lambda - M(\Lambda))|0\right>  \ , \nonumber \\
   &\bar{\lambda}=\left<0|\lambda|0\right> = M(\Lambda)  \ ,
\end{align}
where $\{,\}$ is an anti-commutator. We used the condition (\ref{phys}) for the vacuum state $\left.|0\right>$
since it is included in the physical states. This identity is consistent with the general solution (\ref{so}).
%
If we choose $\bar{\lambda}\neq M(\Lambda)$, the BRS symmetry is spontaneously broken and the BRS charge (\ref{QB}) would not be well-defined.
In this case, the unitarity would be violated since the quartet mechanism does not work. Therefore, we allow the solution $\bar{\lambda}=M(\Lambda)$ only.
%
Then, the potential $U$ becomes
\begin{equation}
\label{ }
U(\bar{\lambda}=M) =  \Lambda \ ,
\end{equation}
and if $\Lambda > 0$, the system has an exact dS solution
%
\begin{align}
\label{dS}
    \bar{\lambda}&= M(\Lambda) , \nonumber \\
    \bar{\phi} &= \phi_0 + \phi_1{\rm e}^{-3Ht} + \frac{u_1}{3H}t \ , \nonumber \\
    H^2 &=\frac{U(\bar{\lambda}=M)}{3M_\text{Pl}^2}=\frac{\Lambda}{3M_\text{Pl}^2} ,\quad a = {\rm e}^{Ht} \ , 
\end{align}
where $\phi_0$ and $\phi_1$ are the constants of integration. $u_1$ becomes a constant since $\bar{\lambda}$ is forced to be $M(\Lambda)$.  
If we take $\bar{\lambda}\neq M(\Lambda)$ or choose $v$ as is satisfied with $v(M)=\infty$, we get $U\neq\Lambda$ and  have a vacuum energy from the potential even if $\Lambda=0$.
We must prohibit, however, 
such a solution since it breaks the BRS symmetry and thus the unitarity or makes the gauge fixing function singular.  

The solution (\ref{dS}) indicates that the BRS sector gives us a new description for dS spacetime. Adding the BRS sector as we did, we have the dS spacetime sustained by not the cosmological constant but the potential energy provided by the vacuum expectation value of $\lambda$. The energy scale of dS spacetime is, however, exactly the same as the original cosmological constant, and we can never have any other scales because of the WT identity (\ref{WT}). In the model, the fine-tuning problems of the cosmological constant are replaced with the fine-tuning for the renormalization of the vacuum expectation value of $\lambda$. Therefore, although the BRS sector can eliminate the cosmological constant from the action, it does not eliminate $\Lambda$ from the evolution of the universe \textit{but just gives another point of view for de Sitter spacetime.} In other words, this is the gauge fixing of dS spacetime by the specific BRS sector. As we will see in the following section, the BRS sector gives a new description also for the quantum field theory in dS spacetime.


Although the BRS sector is excluded from the physical states by Eq. (\ref{phys}),
the field $\phi$ has the non-trivial vacuum expectation value regardless of the value of $M(\Lambda)$ to realize the dS spacetime.
From Eq. (\ref{F}), however, the evolution of $\bar{\phi}$ does not affect the background evolution of the universe as long as
the WT identity (\ref{WT}) is hold. Hence, we can never observe the background $\bar{\phi}$ and the unitarity can be preserved at background level.

To investigate the stability of the solution, we consider a small perturbation $\xi(t)$ around the background solution for $\bar{\phi}(t)$
\begin{align}
\label{}
   \bar{\phi}(t)= \phi_0 + \phi_1{\rm e}^{-3Ht} + \frac{u_1}{3H}t + \xi(t) \ .
\end{align}
We do not need to consider the small perturbations for $\bar{\lambda}$ and $a(t)$ since $\bar{\lambda}$ is fixed to be $M(\Lambda)$ by the WT identity and $a(t)$ is not an independent degree of freedom of the system. Using Eq. \eqref{phi}, we obtain the evolution equation for $\xi(t)$
\begin{align}
\label{}
    \ddot{\xi}+ 3H\dot{\xi}= 0  \ ,
\end{align}
and its solution
\begin{equation}
\label{ }
\xi= \xi_0 + \xi_1\text{exp}(-3Ht)  \ ,
\end{equation}
where $\xi_0$ and $\xi_1$ are the constants of integration.
The small perturbation $\xi$ does not grow, and thus the background solution (\ref{dS}) is stable.


\section{Perturbations around the de Sitter background}\label{sec3}

In the last section, we found that the BRS sector, especially the vacuum expectation value of $\lambda$, gives the new description for de Sitter spacetime. Then, how do perturbations of the BRS sector behave in the de Sitter background? In this section,  
we consider the perturbations of the BRS and gravity sectors around the dS background.
We first confirm that we have the BRS symmetry even to the perturbations around the dS background.
If we do not have the BRS symmetry to the perturbations around the dS background, the BRS symmetry is spontaneously broken.
Such a situation should not occur since we select the background solution which is required by the WT identity (\ref{WT}).

We expand the BRS sector around the dS background and define the perturbations  as
\begin{equation}
\label{BRS2}
\lambda = \bar{\lambda} + \delta\lambda,\quad \phi = \bar{\phi} + \delta \phi, \quad
b = \delta b,\quad c = \delta c \ .
\end{equation}
%
The perturbations interact with the gravity sector nonlinearly.
The background fields $\bar{\lambda}$ and $\bar{\phi}$ are just $c$-numbers, so that they are invariant under the BRS transformation:
$\db \bar{\lambda} = \db \bar{\phi} = 0$.
Then, from Eq. (\ref{BRS}), we obtain the BRS transformation for the perturbations as
\begin{equation}
\label{BRS3}
\db \dl = \db \delta c =0, \quad \db \dph = \delta c,\quad \db \delta b = i\dl \ .
\end{equation}
%
The action expanded around the dS background is given by
\begin{equation}
\label{S3}
S = \int d^4x\sqrt{-g}\left[ \frac{M_\text{Pl}^2}{2}R - U(\bar{\lambda}+\dl) +\dot{\bar{\phi}}g^{0\mu}\partial_\mu\dl
               +g^{\mu\nu}\partial_\mu \dl \partial_\nu \dph
               + i g^{\mu\nu}\partial_\mu \delta b \partial_\nu \delta c\right] \ .
\end{equation}
%
%
The tadpoles will vanish by the background equations (\ref{F})-(\ref{phi}).
The Lagrangian density is invariant under the BRS transformation (\ref{BRS3})
\begin{equation}
\label{ }
\db {\cal L} = \sqrt{-g}\left\{ g^{\mu\nu}
\partial_\mu \dl \partial_\nu \delta c - g^{\mu\nu}\partial_\mu \dl \partial_\nu \delta c\right\} = 0 \ ,
\end{equation}
hence we have the BRS symmetry even to the perturbations around the dS background. In addition,
we have the symmetry under the scale transformation $\delta c\rightarrow {\rm e}^{\rho}\delta c,\ \delta b\rightarrow {\rm e}^{-\rho}\delta b$. The corresponding charges of these symmetries are
\begin{align}
\label{QBP}
    Q_B &= \int d^3x\sqrt{-g}g^{0\mu}\left( \partial_\mu \delta \lambda \delta c - \partial_\mu \delta c\,\delta \lambda\right) \ , \\
    Q_c &= -i\int d^3x\sqrt{-g}g^{0\mu}\left(\delta b\,\partial_\mu \delta c - \partial_\mu \delta b \,\delta c\right)  \ ,
\end{align}
and the charge $Q_c$ assigns the ghost number $1$ for $\delta c$ and $-1$
for $\delta b$. These two charges coincide with the charges (\ref{QB}) and (\ref{QC}) expanded around the dS background.
Thus, the BRS symmetry for the perturbations is intrinsically equivalent to the original BRS symmetry.
%
We note that we have the BRS symmetry before solving the constraints of the gravity sector, but we are not sure if we have the same BRS symmetry after solving the constraints.
As for this point, if we deal with the constraints by adding another set of BRS quartet terms for the gravity sector \cite{Nakanishi:1977gt}, we can preserve the BRS symmetry for ($\delta \phi ,\, \delta \lambda ,\, \delta b,\, \delta c$) sector.
On the other hand, as in the cosmological perturbation theory, if we directly solve the constraints of the gravity sector, it is obscure whether we can preserve the BRS symmetry. The BRS symmetry for ($\delta \phi ,\, \delta \lambda ,\, \delta b,\, \delta c$) sector, however, should not be affected by how to deal with the constraints of the gravity sector. Therefore, \textit{it is plausible to assume} that we can preserve the BRS symmetry even if we solve the constraints directly.
Under this assumption, we will have the BRS charge $Q_B^D$ where the subscript $D$ implies the charge is obtained by solving the constraints directly. Defining the physical states as
\begin{equation}
\label{phys2}
Q_B^{D}\left.|{\rm phys}\right> = 0 \ ,
\end{equation}
we could exclude the negative norm states by the quartet mechanism, and we could safely construct the perturbation theory around the dS background as in the cosmological perturbation theory.
Indeed, as we will show below, at least for the linear perturbations, we have the BRS symmetry after solving the constraints directly.

From now on, we concentrate on the linear perturbations and derive the second-order action.
At the linear level, it is enough to consider only the scalar perturbations since the BRS sector consists of only the scalar perturbations and decouples from the vector and tensor perturbations of the gravity sector.
As for the gravity sector, we thus consider only the scalar perturbations and take the Newtonian gauge \cite{Mukhanov:1990me}
\begin{equation}
\label{ }
ds^2 = -(1+2A)dt^2 + a^2(1-2\psi)d\bo{x}^2 \ .
\end{equation}
We define the linear perturbations of the BRS sector as similarly as in (\ref{BRS2}), but we ignore the nonlinear interaction with the gravity in the linear analysis. We expand the potential $U$ up to the second order of perturbation as
\begin{equation}
\label{ }
U(\bar{\lambda} + \dl) = u_1\dl + u_2\dl^2 \ , 
\end{equation}
where $u_1$ was defined in (\ref{phi}) and $u_2\equiv \left.\frac{1}{2}\frac{d^2U}{d\lambda^2}\right|_{\lambda = M}$.
%
%
The full second-order action in the Newtonian gauge is derived straightforwardly as
\begin{align}
\label{S2}
S_2^{(S)}= \int dtd^3x a^3&\left[ 3M_\text{Pl}^2(-\dot{\psi}^2- 2HA\dot{\psi}-H^2A^2) + M_\text{Pl}^2a^{-2}(\psi_{,i}^2 +2A\partial^2\psi)\right. \nonumber \\
                &  +\dot{\bar{\phi}}(A+3\psi)\dot{\dl} -u_1(A-3\psi)\dl - u_2\dl^2 \nonumber \\
            &\left.    -\left( \dot{\dl}\dot{\dph} - a^{-2}\dl_{,i}\dph_{,i}\right)-i\left( \dot{\delta b}\dot{\delta c} - a^{-2}\delta b_{,i}\delta c_{,i}\right) \right] \ ,
\end{align}
where the Latin indices denote the spatial derivatives and $\partial^2 \equiv \partial_i\partial_i$.
The Newtonian potential $A$ is an auxiliary field, and its equation of motion yields the Hamiltonian constraint
\begin{equation}
\label{A}
A = -\frac{\dot{\psi}}{H} +\frac{a^{-2}\partial^2\psi}{3H^2} + \frac{\dot{\bar{\phi}}}{6M_\text{Pl}^2H^2}\dot{\dl} -\frac{u_1}{6M_\text{Pl}^2H^2}\dl \ .
\end{equation}
We define the gauge invariant Mukhanov-Sasaki variable \cite{Sasaki:1986hm, Mukhanov:1988jd},
\begin{equation}
\label{MS}
\dpt = \delta \phi + \frac{\dot{\bar{\phi}}}{H}\psi \ .
\end{equation}
The other scalar perturbations of the BRS sector are gauge invariant in themselves since all the time derivatives of their background fields vanish.
Using the constraint (\ref{A}) and the new variable (\ref{MS}), the action (\ref{S2}) is reduced to
\begin{align}
\label{ }
S_2^{(S)}= \int dtd^3xa^3&\left[ \frac{M_\text{Pl}^2a^{-4}}{3H^2}(\partial^2\psi)^2
      + \frac{a^{-2}}{H}\left(\dot{\bar{\phi}}- \frac{u_1}{3H}\right)\dl \partial^2\psi
      +\frac{a^{-2}\dot{\bar{\phi}}}{3H^2}\dot{\dl}\partial^2 \psi \right. \nonumber \\
    &+\frac{1}{M_{\rm Pl}^2}\left(\frac{\dot{\bar{\phi}}^2}{12H^2}\dot{\dl}^2 + \frac{u_1^2}{6H^2}\dl^2\right) - u_2\dl^2 \nonumber \\
    &\left.-\left( \dot{\dl}\dot{\dpt} - a^{-2}\dl_{,i}\dpt_{,i}\right) 
          -i\left( \dot{\delta b}\dot{\delta c} - a^{-2}\delta b_{,i}\delta c_{,i}\right)\right] \ .
\end{align}
The curvature perturbation $\psi$ also becomes an auxiliary field, so that its equation of motion yields another constraint
\begin{equation}
\label{mc}
\partial^2 \psi = \frac{a^2}{2M_{\rm Pl}^2}\left((u_1-3H\dot{\bar{\phi}})\dl - \dot{\bar{\phi}}\dot{\dl}\right) \ .
\end{equation}
Using the constraint (\ref{mc}) and the background solution of $\bar{\phi}$ in (\ref{dS}), we obtain the second-order action for the dynamical fields
\begin{align}
\label{SG}
S_2^{(S)}  =\int dtd^3x&a^3 \left[  -\left( \dot{\dl}\dot{\dpt} - a^{-2}\dl_{,i}\dpt_{,i}\right)
           -i\left( \dot{\delta b}\dot{\delta c}- a^{-2}\delta b_{,i}\delta c_{,i}\right)\right. \nonumber \\
            &\quad + \left.\left(\frac{u_1^2}{6M_{\rm Pl}^2H^2}-u_2 -\frac{27H^2}{2M_{\rm Pl}^2}\phi_1^2a^{-6}\right)\dl^2\right]\ .
\end{align}
This action is gauge invariant although we started from the Newtonian gauge, and corresponds to the result in Ref.~\refcite{Langlois:2008mn} when taking the flat gauge $\psi=0$. The BRS transformation for the linear perturbations is given as the same as (\ref{BRS3}).
The curvature perturbation is invariant under the BRS transformation, $\db \psi =0$, so that $\db \dpt = \db \delta c$.
%
Then, we easily find the action (\ref{SG}) is invariant under the BRS transformation
\begin{equation}
\label{BS2}
\db S_2^{(S)} = \int dtd^3xa^3\left[ -\left( \dot{\dl}\dot{\delta c} - a^{-2}\dl_{,i}\delta c_{,i}\right) + \left( \dot{\dl}\dot{\delta c} - a^{-2}\dl_{,i}\delta c_{,i}\right)\right] = 0 \ .
\end{equation}
Therefore, at least to the linear perturbation, we actually have the BRS symmetry after solving constraints.
\section{Quantum fluctuations}\label{sec4}

In this section, we consider the quantization of the linear perturbations and calculate the two point functions of them following the usual method of cosmological perturbation theory.

\subsection{Quantization of linear perturbations}

We introduce new variables
\begin{equation}
\label{ }
\ul = a\dl,\quad \up = a\dpt, \quad u_b = a\delta b,\quad u_c = a\delta c \ ,
\end{equation}
and the conformal time
\begin{equation}
\label{ }
\eta = \int^t \frac{dt'}{a} \ .
\end{equation}
Using them, we can rewrite the action (\ref{SG})  as
\begin{align}
\label{SE}
S_2^{(S)} &= \int d\eta d^3x {\cal L}_2^{(S)} \nonumber \\ 
&= \int d\eta d^3x\left[  -\left( \ul'\up' - {\ul}_{,i}{\up}_{,i} +\frac{a''}{a}\ul\up\right) 
          -i\left( u_b'u_c' - u_{b,i}u_{c,i}+\frac{a''}{a}u_bu_c\right)\right. \nonumber \\
          &\qquad\qquad\quad \left. +\left\{\left(\frac{u_1^2}{6M_\text{Pl}^2H^2}-u_2\right)a^2-\frac{27H^2}{2M_{\rm Pl}^2}\phi_1^2a^{-4}\right\}\ul^2\right] \ ,
\end{align}
where the prime denotes the derivative with respect to the conformal time $\eta$.
The BRS transformations (\ref{BRS3}) for the new variables become
\begin{equation}
\label{BRS4}
\db \ul = \db u_c = 0,\quad \db \up = u_c,\quad \db u_b = i\ul \ .
\end{equation}
The conjugate momenta are given by
\begin{align}
\label{}
    \pl&= \partial {\cal L}_2^{(S)}/\partial \ul' = -\up' \ ,  \nonumber \\
    \pp&= \partial {\cal L}_2^{(S)}/\partial \up' = -\ul' \ , \nonumber \\
    \pb&= \partial {\cal L}_2^{(S)}/\partial u_b' = iu_c'  \ , \nonumber\\
    \pc&= \partial {\cal L}_2^{(S)}/\partial u_c' = -iu_b'  \ .
\end{align}
We used the right-hand derivative defined as follows
\begin{align}
\label{ }
&\partial (AB)/\partial\theta = A(\partial B/\partial\theta) + (-)^{|B|}(\partial A/\partial\theta)B \ ,
\end{align}
where $\theta$ is a Grassmann number and $|B|$ is the statistical index of $B$.
We perform the canonical quantization in the usual way
\begin{align}
\label{cr}
    [\ul(\bo{x}, \eta),\,\pl(\bo{y},\eta) ]&= i\delta^3(\bo{x}-\bo{y})  \ , \nonumber \\
    [\up(\bo{x}, \eta),\,\pp(\bo{y},\eta) ]&= i\delta^3(\bo{x}-\bo{y})   \ ,\nonumber \\
    \{u_b(\bo{x}, \eta),\,\pb(\bo{y},\eta) \}&= i\delta^3(\bo{x}-\bo{y}) \ ,  \nonumber \\
    \{u_c(\bo{x}, \eta),\,\pc(\bo{y},\eta) \}&= i\delta^3(\bo{x}-\bo{y})   \ ,
\end{align}
where $[,]$ is a commutator. Conserved charges corresponding to the BRS symmetry and the scale transformation $u_c\rightarrow {\rm e}^{\rho}u_c,\ u_b\rightarrow {\rm e}^{-\rho}u_b$ are respectively given by
\begin{align}
\label{ }
\delta Q_B &= \int d^3x (-\ul' u_c + u_c'\ul) = \int d^3x (\pp u_c -i\pb\ul)\ , \\
\delta Q_c &= i\int d^3x(u_bu_c'-u_b'u_c) = \int d^3x (u_b\pb+\pc u_c) \ .
\end{align}
We can easily check that $\delta Q_B$ is the generator of the BRS transformation (\ref{BRS4}), and $\delta Q_c$ assigns the ghost number $1$ for $u_c$ and $-1$ for $u_b$.
%
%
%
In the last section, we assume that we have the BRS symmetry and the corresponding charge $Q_B^{D}$ even if we solve the gravity constraints directly.
We can interpret the charge $\delta Q_B$ as the asymptotic part of the charge $Q_B^D$ apart from the renormalization constant.
The fields $u_\phi$ and $u_I$ behave as the asymptotic fields apart from the renormalization constant.
Thus, we obtain the following relations
\begin{align}
\label{asym}
&[iQ_B^D,\, \up] = [i\delta Q_B,\, \up] = u_c\ , \nonumber  \\
&[iQ_B^D,\, \ul] = [i\delta Q_B,\, \ul]= 0 \ , \nonumber  \\
&\{iQ_B^D,\, u_b\} = \{i\delta Q_B,\, u_b\}=i\ul \ , \nonumber  \\
&\{iQ_B^D,\, u_c\}= \{i\delta Q_B,\, u_c\} = 0 \ .
\end{align}
See Appendix for the proof of these relations.

Then, we expand the linear perturbations by using creation and annihilation operators, and investigate the metric matrix of the Fock space of the linear perturbations.
The equations of motion for the linear perturbations are derived from (\ref{SE}) as
\begin{align}
\label{eom2}
    &\left( \partial_\eta^2 - \partial^2 + \frac{a''}{a}\right) u_I(x) = 0, \quad I=(\lambda, b, c)\ , \\
    \label{eom3}
    &\left( \partial_\eta^2 - \partial^2 + \frac{a''}{a}\right) \up(x) = -r\ul(x) \ , \nonumber \\
     &\quad r\equiv
    \left(\frac{u_1^2}{3M_\text{Pl}^2H^2}-2u_2\right)a^2  - \frac{27H^2}{M_{\rm Pl}^2}\phi_1^2a^{-4} \ ,
\end{align}
where $\partial_\eta^2\equiv \partial^2/\partial \eta^2$. The fields $u_I(x)$ satisfy the equations of motion for free scalar fields, so we can expand them in a usual way using the creation and annihilation operators $I_\bo{k}^{(\dag)}$ as
\begin{align}
\label{}
    u_I(x)= \int \frac{d^3k}{(2\pi)^{3/2}}\left\{ I_\bo{k} f_k(\eta) {\rm e}^{i\bo{k}\cdot\bo{x}} +
    I_\bo{k}^\dag f_k^*(\eta) {\rm e}^{-i\bo{k}\cdot\bo{x}}
    \right\}, \quad k\equiv |\bo{k}| \ ,
\end{align}
where $f_k(\eta)$ is the mode function for $u_I(x)$. The mode function satisfies
\begin{equation}
\label{ }
f_k'' + \left( k^2 - \frac{2}{\eta^2}\right)f_k = 0 \ .
\end{equation}
Demanding that the solution on small scales behaves like the Minkowski vacuum, we obtain the solution for the positive frequency modes \cite{Bunch:1978yq}
\begin{equation}
\label{ }
f_k(\eta) = \frac{1}{\sqrt{2k}}\left( 1 -\frac{i}{k\eta}\right) {\rm e}^{-ik\eta} \ ,
\end{equation}
and its complex conjugate for the negative frequency modes as the other solution. Note that for the normalization of (anti-)commutation relations of $I_\bo{k}^{(\dag)}$, we imposed the following condition
\begin{equation}
\label{ }
f_kf_k^{*'} - f_k'f_k^* = i \ .
\end{equation}
On the other hand, the field $\up(x)$ are satisfied with the linear differential equation with a source.
Performing the Fourier transformation
\begin{align}
\label{}
    \up(x) = \int  \frac{d^3k}{(2\pi)^{3/2}}\hat{u}_{\phi\bo{k}}(\eta){\rm e}^{i\bo{k}\cdot\bo{x}}  \ ,
\end{align}
we obtain the linear differential equation for $\hat{u}_{\phi\bo{k}}$
\begin{equation}
\label{ueq}
\hat{u}_{\phi\bo{k}}'' + \left( k^2 - \frac{2}{\eta^2}\right)\hat{u}_{\phi\bo{k}} = -r(\lambda_\bo{k}f_k + \lambda^\dag_\bo{-k}f_k^*) \ .
\end{equation}
For the homogeneous equation
\begin{equation}
\label{ }
\hat{u}_{\phi\bo{k}}^{(0)''} + \left( k^2 - \frac{2}{\eta^2}\right)\hat{u}_{\phi\bo{k}}^{(0)} = 0 \ ,
\end{equation}
we obtain the solution
\begin{equation}
\label{ }
\hat{u}_{\phi\bo{k}}^{(0)}= \phi_\bo{k}f_k + \phi^\dag_\bo{-k}f_k^* \ ,
\end{equation}
where $\phi_\bo{k}$ and $\phi^\dag_\bo{-k}$ are the creation and annihilation operators for the field $\phi$.
A special solution to Eq. (\ref{ueq}) is given by
\begin{align}
\label{}
    \hat{u}_{\phi\bo{k}}^{(s)} = f_k\int \frac{-Rf_k^*}{W}d\eta + f_k^*\int \frac{Rf_k}{W}d\eta, \quad
    R\equiv  -r(\lambda_\bo{k}f_k + \lambda^\dag_\bo{-k}f_k^*) \ ,
\end{align}
where the Wronskian $W$ is $W = f_kf_k^{*'} - f_k'f_k^* = i$.
The special solution can be decomposed as
\begin{align}
\label{}
    \hat{u}_{\phi\bo{k}}^{(s)}& =\lambda_\bo{k}P_k(\eta) + \lambda^\dag_\bo{-k}P_k^*(\eta)  \ , \\
    P_k(\eta)&= -i\left( f_k\int r|f_k|^2d\eta -f_k^*\int rf_k^2d\eta\right)   \ . \nonumber
\end{align}
The complete solution to Eq. (\ref{ueq}) is given by the sum of the homogeneous solution and the special solution $\hat{u}_{\phi\bo{k}}=  \hat{u}_{\phi\bo{k}}^{(0)} + \hat{u}_{\phi\bo{k}}^{(s)}$, so the field $\up(x)$ expands as
\begin{equation}
\label{up}
\up(x) =  \int \frac{d^3k}{(2\pi)^{3/2}}\left\{\left( \phi_\bo{k} f_k +\lambda_\bo{k}P_k \right){\rm e}^{i\bo{k}\cdot\bo{x}} +
    \left(\phi_\bo{k}^\dag f_k^* + \lambda_\bo{k}^\dag P_k^*\right) {\rm e}^{-i\bo{k}\cdot\bo{x}}
    \right\} \ .
\end{equation}

We get the following relations for the creation operators
\begin{align}
\label{I}
I_\bo{k} &= i\int \frac{d^3x}{(2\pi)^{3/2}}{\rm e}^{-i\bo{k}\cdot\bo{x}}(f_k^*u_I'(x) - f_k^{*'}u_I(x)) \ ,\\
\label{ph}
\phi_\bo{k} &= i\int \frac{d^3x}{(2\pi)^{3/2}}{\rm e}^{-i\bo{k}\cdot\bo{x}}(f_k^*u_\phi'(x) - f_k^{*'}\up(x)) \nonumber \\
      &\quad-i \left\{ \lambda_\bo{k}(f_k^*P_k'-f_k^{*'}P_k) + \lambda_\bo{-k}^\dag(f_k^*P_k^{*'} - f_k^{*'}P_k^*)\right\} \ .
\end{align}
Using (\ref{I}), (\ref{ph}) and the (anti-)commutation relations (\ref{cr}), we obtain the (anti-)commutation relations for the creation and annihilation operators
\begin{align}
\label{}
    [\phi_\bo{k},\, \phi_\bo{q}^\dag] &= [\phi_\bo{k},\, \phi_\bo{q}] = 0 \ , \nonumber \\
    [\phi_\bo{k},\, \lambda_\bo{q}^\dag]&=[\lambda_\bo{k},\, \phi_\bo{q}^\dag]= -\delta^3(\bo{k}-\bo{q})  \ , \nonumber \\
    \{b_\bo{k},\,c_\bo{q}^\dag\}&= -\{c_\bo{k},\,b_\bo{q}^\dag\} = -i\delta^3(\bo{k}-\bo{q}) \ ,
\end{align}
and all the other (anti-)commutation relations become zero. The metric matrix of the Fock space is
\begin{align}
\label{metric}
\eta_{kq} = \bordermatrix{
                       & \phi_\bo{q}^\dag & \lambda_\bo{q}^\dag & b_\bo{q}^\dag & c_\bo{q}^\dag \cr
     \phi_\bo{k} &  0                        & -\delta_{kq} & 0                 & 0                    \cr
     \lambda_\bo{k} & -\delta_{kq} & 0                   & 0                 & 0                    \cr
     b_\bo{k}     & 0                         & 0                     & 0        & -i\delta_{kq}       \cr
     c_\bo{k}     & 0                         & 0                     & i\delta^3_{kq}  & 0
     }, \quad
      \delta_{kq} \equiv \delta^3(\bo{k}-\bo{q})\ .
\end{align}
From the structure of metric matrix, we find that there exist the negative norm states in the Fock space. However,
defining the physical states by Eq. (\ref{phys2}), we could exclude the negative norm states from the physical Fock space by the quartet mechanism.

\subsection{Two point correlators}

%
%
%
Next, we consider the two point correlators of the linear perturbations. We introduce the following perturbations \cite{Bassett:2005xm}
\begin{equation}
\label{ }
{\cal R} \equiv \frac{H}{\dot{\bar{\phi}}}\widetilde{\delta \phi},\quad {\cal S}_I \equiv \frac{H}{\dot{\bar{\phi}}}\delta I, \quad I = (\lambda, b, c) \ ,
\end{equation}
and calculate the two point correlators of them.
${\cal R}$ is the gauge invariant curvature perturbation or the adiabatic perturbation since $\phi$ is the adiabatic field \cite{Gordon:2000hv}.
The other modes ${\cal S}_I$ are regarded as the isocurvature perturbations.
We define the two point correlation function in the Fourier space as
\begin{equation}
\label{ }
\left<0| X(\bo{x}, t)Y(\bo{y}, t) |0\right> \equiv \int \frac{d^3k}{(2\pi)^3}\frac{2\pi^2}{k^3}{\cal C}_{XY}{\rm e}^{i\bo{k}\cdot(\bo{x}-\bo{y})} \ ,
\end{equation}
where $X$ and $Y$ are arbitrary fields. We are interested in the two point correlators of the linear perturbations, so we identify the vacuum state $\left. |0\right>$ as the free vacuum.
Using Eqs. (\ref{up}) and (\ref{metric}), we obtain 
\begin{align}
\label{}
    {\cal C}_{\up\up}& = -\frac{k^3}{2\pi^2}(f_k^*P_k + f_kP_k^*)= -\frac{k^3}{2\pi^2}2{\rm Im}\left( f_k^2\int rf_k^{*2}d\eta\right) \ , \nonumber \\
    {\rm Im}\left( f_k^2\int rf_k^{*2}d\eta\right)&= \frac{1}{4k^3H^2\eta^2}\left[ \left(\frac{u_1^2}{3M_{\rm Pl}^2H^2}-2u_2\right)\left(\frac{2}{3}{\rm cos}(2k\eta){\rm ln}|\eta| + C\right) \right. \nonumber \\
   &\qquad \qquad\qquad \left. -\frac{27\cdot 7H^2}{4M_{\rm Pl}^2}\phi_1^2\left(\frac{H}{k}\right)^6
     + O(k\eta)\right]  \ ,
\end{align}
where $C$ is the constant of integration and $O(k\eta)$ is the contribution from sub horizon scales which can be ignored on super horizon scales.
Then, we obtain the power spectrum of ${\cal R}$,
\begin{align}
\label{}
   {\cal P}_{\cal R} &\equiv {\cal C}_{\cal RR} = \left(\frac{H}{a\dot{\bar{\phi}}}\right)^2{\cal C}_{\up\up}   \nonumber \\
      &=  \frac{H^2}{2\pi^2(u_1-9H^2\phi_1{\rm e}^{-3Ht})^2}\left\{
      \left(\frac{u_1^2}{M_{\rm Pl}^2}-6H^2u_2\right)(Ht + C) \right. \nonumber \\
      &\qquad\qquad\qquad\qquad\qquad\qquad\left.+ \frac{27\cdot 63}{8}\frac{H^4\phi_1^2}{M_{\rm Pl}^2}\left(\frac{H}{k}\right)^6
      + O(k\eta)\right\} \ ,
\end{align}
where we suitably redefined the constant of integration $C$.
On the other hand, since all the diagonal parts of the metric matrix (\ref{metric}) are zero, the power spectra of isocurvature perturbations vanish on all scales
\begin{equation}
\label{ }
{\cal P}_{{\cal S}_I} \equiv {\cal C}_{{\cal S}_I{\cal S}_I}\propto \int d^3q\left< 0|I_\bo{k}I_\bo{q}^\dag|0\right> = 0 \ .
\end{equation}
The nonzero cross correlations are given by
\begin{align}
\label{cr7}
&{\cal C}_{\widetilde{\delta\phi}\delta\lambda} = i\,{\cal C}_{\delta c\delta b}= -\frac{k^3}{2\pi^2}\left|\frac{f_k}{a}\right|^2=
-\left(\frac{H}{2\pi}\right)^2 + O((k\eta)^2)  \ , \nonumber \\
&{\cal C}_{{\cal R}{\cal S}_\lambda} = i\,{\cal C}_{{\cal S}_c{\cal S}_b}=
-\frac{9H^4}{(u_1-9H^2\phi_1{\rm e}^{-3Ht})^2}\left(\frac{H}{2\pi}\right)^2 + O((k\eta)^2) \ .
\end{align}
We can obtain the relation ${\cal C}_{\widetilde{\delta\phi}\delta\lambda} = i\,{\cal C}_{\delta c\delta b}$ as a result of the WT identity for the linear perturbations and the free vacuum
\begin{align}
\label{}
    &0= \left<0|[i Q_B^D,\,\up(\bo{x}, \eta)u_b(\bo{y},\eta)]|0\right> \nonumber \\
    &\ \,=\left<0|[i\delta Q_B,\,\up(\bo{x}, \eta)u_b(\bo{y},\eta)]|0\right> \nonumber \\
     &\ \,= \left<0|u_c(\bo{x}, \eta)u_b(\bo{y},\eta)|0\right> + i\left<0|\up(\bo{x}, \eta)\ul(\bo{y},\eta)]|0\right> \ ,   \nonumber  \\
    &\left<0|\up(\bo{x}, \eta)\ul(\bo{y},\eta)]|0\right> = i\left<0|u_c(\bo{x}, \eta)u_b(\bo{y},\eta)|0\right> \ ,
\end{align}
where we used the relations (\ref{asym}) for the second equality.

We obtained nonzero two point functions of the scalar perturbations although the classical system without the BRS sector has no dynamical scalar mode. 
We can treat the BRS sector as multi scalar fields on the dS spacetime and obtain the two point functions among them as in the multi-field inflation models \cite{Polarski:1994rz, Gordon:2000hv, Amendola:2001ni, Langlois:2008mn, Langlois:2008qf}.
The power spectrum ${\cal P}_{\cal R}$ sourced by the isocurvature perturbation ${\cal S}_\lambda$ 
grows on super horizon scales and diverges in the infrared limit $k=0$ when $\phi_1\neq0$.
However, this does not imply that we can not add the BRS sector. All of the two point functions we derived are not BRS-invariant, and thus they are never observed even if they diverge.
Instead, 
they give a new description for the quantum field theory in dS spacetime. When considering the perturbation theory of the physical sector like graviton, the BRS sector can contribute to the loop integrals. It would give a new interpretation to the physical quantities in dS spacetime 
\footnote{There is a work to solve the infrared divergence in the two point function of a scalar field on $S^4$ by adding a BRS sector\cite{Folacci:1992xc}.}. Deriving the two point functions of the BRS sector, we constructed a basis of the perturbation theory with the BRS sector in dS spacetime obeying the in-in formalism.

\section{Conclusion}\label{sec5}

We proposed the topological model with the generic potential of scalar field $\lambda$, which was constructed by using
the BRS quartet terms as gauge fixings for the Lagrangian density ${\cal L}_\phi=0$,
with the Einstein-Hilbert term and the cosmological constant $\Lambda$. 
We also discussed the perturbations around the dS background solution and the two point correlators.
After we shortly review the topological model proposed in Ref.~\refcite{Nojiri:2016mlb}, we generalized the model so as to have the generic potential of $\lambda$, and eliminated the cosmological constant $\Lambda$ from the action by using the potential.
Then, we constructed the dS background solution with the BRS sector.
We found that we can select the dS solution without spontaneously breaking the BRS symmetry by imposing the WT identity.
The dS solution is sustained by not the cosmological constant but the potential energy provided by the vacuum expectation value of $\lambda$. The energy scale of dS spacetime is, however, exactly the same as the eliminated cosmological constant. This exhibits truly that the BRS sector does not eliminate $\Lambda$ from the evolution of the universe but gives another point of view for dS spacetime, say, a gauge fixing of dS spacetime. In the model, the fine-tuning problems of the cosmological constant is replaced to the fine-tuning for the renormalization of the vacuum expectation value of $\lambda$.
Although one of the BRS sector, $\phi$, also has a nonzero background value, it can not be observed since it does not concern with the background evolution of the universe as long as the WT identity is hold.
We studied the stability of the dS solution and confirmed that the solution is stable.

We studied the BRS symmetry for the perturbations around the dS background.
We have the same BRS symmetry as the original one for the perturbations around the dS background before solving the gravitational constraints
since we take the background without the spontaneous symmetry breaking.
The BRS symmetry for the perturbations will be preserved even if we deal with the constraints by adding other BRS quartet terms for the gravitational sector.
Based on the argument that the BRS symmetry should not be affected by how the constraints are dealt with,
we naturally assume that we have the BRS symmetry even if we solve the constraints directly.
Under this assumption, we will have the BRS charge $Q_B^D$.
Defining the physical states as annihilated by $Q_B^D$, we could eliminate the negative norm states by Kugo-Ojima's quartet mechanism and safely construct the perturbation theory around the dS background.
We confirmed that at least for the linear perturbations, we have the BRS symmetry even after solving the constraints directly.

Using the BRS quartet terms, we constructed the cosmological perturbation theory around the dS background. Following the usual procedure,
we quantized the linear perturbations and calculated the two point functions of the gauge invariant curvature perturbation and the isocurvature perturbations due to the BRS sector.
Although we have no dynamical scalar mode in the classical theory without the BRS sector,
we obtained the power spectrum of the curvature perturbation and the cross correlations of the curvature perturbation and the isocurvature perturbation, but we have no power spectrum for the isocurvature perturbations.
The power spectrum of curvature perturbation sourced by one of the isocurvature perturbations can grow even on super horizon scales and diverge in the infrared limit when one of the initial condition is satisfied with $\phi_1\neq0$. 
This, however, does not imply that we can not define the perturbation theory if we add the BRS sector. We do never observe the nonzero two point functions since all of them are not BRS-invariant.
Nevertheless, it would be worth to consider the BRS sector since it would give a new description for the quantum field theory in dS spacetime. When considering the perturbation theory, the BRS sector contributes to the loop integrals of the physical sector such as graviton. It would give a new interpretation to the physical quantities in dS spacetime.
In this work, deriving the dS solution and the two point functions of the BRS sector, we give a basis for a new description of dS spacetime and the perturbation theory with the BRS sector in dS spacetime. We believe that this would be important for the better understanding of the quantum field theory in dS spacetime and deserve further investigation.

\section*{Acknowledgment}

R.S. thanks to S. Nojiri for useful discussions.
This work was supported partially by the National Natural Science Foundation of China under Grant No. 11475065 and the Major Program of the National Natural Science Foundation of China under Grant No. 11690021.


%

\appendix

\section{Proof of the relations (\ref{asym})}

The BRS transformations by $Q_B^D$ for the linear perturbations are
\begin{equation}
\label{ }
[iQ_B^D,\, u_{\alpha}(x)\} = \delta \Phi_\alpha(x) ,\quad \alpha = (\phi, \, I) \ ,
\end{equation}
where $[,\}$ acts as the commutator for Grassmann even quantities and as the anti-commutator for Grassmann odd quantities, $\delta \Phi_\alpha$ is the variation of the linear perturbations $u_\alpha$.
$u_\alpha(x)$ satisfy the equations of motion (\ref{eom2}) and (\ref{eom3}), so for $u_I(x)$, we get
\begin{align}
\label{ }
&[iQ_B^D,\, \Box_\eta u_I(x)\} = \Box_\eta\delta \Phi_I(x) = 0,\quad \Box_\eta \equiv \partial_\eta^2 -\partial^2_{\bo{x}} +\frac{a''}{a}\ , \nonumber \\
&\delta \Phi_I = a_{IJ}u_J \ ,
\end{align}
where $a_{IJ}$ is a constant matrix. Then, we obtain
\begin{equation}
\label{vui}
[iQ_B^D,\, u_I(x)\} = a_{IJ}u_J(x) \ .
\end{equation}
As for $\up(x)$, we get
\begin{align}
\label{}
    &[iQ_B^D,\, \Box_\eta\up(x)] = \left[iQ_B^D,\, -r\ul(x)\right] = -ra_{\lambda J}u_J = \Box_\eta \delta \Phi_\phi(x) \ ,\nonumber   \\
    &\delta \Phi_\phi = a_\alpha u_\alpha \ ,
\end{align}
where we used (\ref{vui}) for the second equality and $a_\alpha$ is a constant vector. Then, we obtain
\begin{equation}
\label{vup}
[iQ_B^D,\, \up(x)]  = a_\alpha u_\alpha(x) \ .
\end{equation}
The BRS charge $Q_B^D$ should be expanded as
\begin{equation}
\label{ }
Q_B^D = \delta Q_B + \cdots \ ,
\end{equation}
where the ellipsis denotes the third and higher order terms of the perturbations.
Only the lowest order term $\delta Q_B$ can generate the linear perturbations on the right hand sides of (\ref{vui}) and (\ref{vup}). Therefore,
we obtain
\begin{equation}
\label{ }
[iQ_B^D,\, u_\alpha(x)\} = [i\delta Q_B,\, u_\alpha(x)\}\ ,
\end{equation}
and the relations (\ref{asym}) are proved.

\end{document}